\documentclass[12pt]{iopart}

\usepackage{iopams}
\usepackage{amsthm}
\usepackage{graphicx}

\makeatletter
\newcommand{\tpitchfork}{%
  \vbox{
    \baselineskip\z@skip
    \lineskip-.52ex
    \lineskiplimit\maxdimen
    \m@th
    \ialign{##\crcr\hidewidth\smash{$-$}\hidewidth\crcr$\pitchfork$\crcr}
  }%
}
\makeatother

\usepackage{color}

\begin{document}
\title{From disformal electrodynamics to exotic spacetime singularities}


\author{Eduardo Bittencourt${}^{1}$, Ricardo Fernandes${}^{1}$, \'Erico Goulart${}^{2}$ and José Eloy Ottoni${}^{2}$}
\address{${}^{1}$Federal University of Itajub\'a, Itajub\'a, Minas Gerais 37500-903, Brazil}
\address{${}^{2}$Federal University of S\~ao Jo\~ao d'el-Rei, C.A.P. Rod.: MG 443, KM 7, CEP-36420-000, Ouro Branco, MG, Brazil}
\ead{bittencourt@unifei.edu.br, ricardofernandes@unifei.edu.br, egoulart@ufsj.edu.br, jeottoni@ufsj.edu.br}

\begin{abstract}
We study different types of spacetime singularities which emerge in the context of disformal electrodynamics. The latter is characterized by transformations of the background metric which preserve regular (non-null) solutions of Maxwell equations in vacuum. Restricting ourselves to the case of electrostatic fields created by charged point particles along a line, we show that exotic types of singularities arise.
\end{abstract}

\vspace{2pc}

%
%
%

\section{Introduction}

One of the most striking features of Einstein's theory of general relativity (GR) is the prediction of singularities. Roughly speaking, in the vicinity of a singularity the spacetime manifold presents undesirable mathematical pathologies, such as the divergence of a physical quantity, the blow-up of a geometrical invariant or the abrupt interruption of the worldline of an observer. The fact that these catastrophic behaviors are commonplace in GR is guaranteed by the so-called singularity theorems, originally due to Penrose, Hawking and others \cite{Penrose,Hawking,hawking1970singularities,hawking2023large}. In general, the architecture behind these theorems (and their generalizations) can be sketched as follows: if a spacetime of sufficient differentiability satisfies some  conditions on the curvature and causality plus an appropriate initial and/or boundary condition, then it contains endless but incomplete causal geodesics. However, as stressed in the references \cite{senovilla1998singularity, senovilla2011singularity, senovilla20151965}, the weakest point of the singularity theorems is precisely their conclusions: the theorems do not say anything, in general, about the position, strength, extension, dimension, shape, topology
and character of the singularity. All one knows is that there exists, at least, one incomplete causal geodesic. What is more: since Einstein's field equations are highly nonlinear, it is not always easy to design generic singular solutions and interpret them adequately (see \cite{ellis1977singular} for a possible classification scheme). Therefore, in order to shed some light on the debate, new ways to construct spacetime singularities are very welcome.

In this paper, we investigate exotic types of singularities that might emerge in the context of \textit{disformal electrodynamics}\footnote{To the best of our knowledge, there is no consensus on the terminology of singularities in a four-dimensional spacetime. Throughout, by `exotic singularity' we mean a singular configuration of the geometry which is not commonplace in ordinary solutions of GR, such as black holes or FRLW geometries.}. More generally, the concept of a disformal transformation has its roots in the early days of GR and is related to a replacement of the underlying causal structure. A simple example was first obtained by W. Gordon in the study of light propagation inside a dielectric medium \cite{Gordon}. Another realization occurs in the so-called Kerr-Schild ansatz, where the Minkowski metric is modified by the inclusion of a geodesic and shear-free light-like vector field \cite{kerr-schild}. In the nineties, Bekenstein proposed a new theory of gravity in the context of Finsler manifolds, with the metrics somehow related via the gradient of a scalar field \cite{beken1,beken2}. After that, this procedure has been investigated in a variety of subjects, ranging from the analogue models of gravity to geometric-inspired modifications of GR \cite{Nov_Vis_Vol,Barcelo2005,beken_mond04,magueijo04,Kaloper04,milgrom09,clifton12,mota12,Novello12,Novello13,scalartheory13,dario13,Bit14,Nov_Bit14,rua14,sakstein14,Arroja15,Ip15,Sakstein15,Yuan15,Carvalho16}. More recently, some of us have demonstrated that the Klein-Gordon equation for the scalar field and the Maxwell equations for the electromagnetic field remain unchanged when subjected to disformal transformations, encompassing the conformal ones \cite{Falciano12,bitt15,Goulart13,Goulart21}. As a larger class of symmetry of well-known field equations for fundamental fields, the scrutiny of all its properties is mandatory.

As discussed in the references \cite{Goulart13,Goulart21}, the case of disformal electrodynamics is characterized by a simple transformation of the spacetime metric which preserves regular (non-null) solutions of Maxwell equations in vacuum i.e., without local currents. The main feature of the transformation is that it gives rise to a new metric tensor with a hyperbolic signature (the disformal metric), which depends on the electromagnetic field non-linearly, and exchanges the roles of space and time in a nontrivial fashion. As noticed in the reference \cite{harte2017metric}, this generalizes the well-known conformal invariance of electrodynamics and provides a sense in which electromagnetism alone cannot be used to measure certain aspects of geometry. Even more dramatically, a congruence of observers that is physically acceptable with respect to the physical metric can become unacceptably spacelike with respect to the disformal metric. This drastic modification of the causal structure is essentially due to algebraic reasons and is related to well-known identities satisfied by the energy-momentum tensor associated with any regular solution\footnote{The reader is referred to \cite{Falciano12,bitt15} for similar constructions in the cases of linear scalar and spinor fields.}.

Our aim here is two-folded: first is to better understand the types of singularities that can emerge in the context of disformal electrodynamics using curvature invariants. It is our hope that it will be easier to understand the true meaning of disformal electrodynamics if one has in mind a clear picture of the local kinds of behavior of singularities and how they relate to global aspects. Secondly, it is to interpret these singularities in terms of sources and sinks of the disformal time. We think that a clear distinction between these types of singularities may make it easier to predict how and why they emerge in GR and avoid them, if necessary. Although the invariance is still valid in a curved background, we shall work within the context of Minkowski spacetime to keep things as simple as possible. Another two simplifying assumptions will be the following: i) the Faraday tensor contains only static electric fields, with charge particles as sources; ii) all particles are co-linear for all space-like sections. Then, removing the worldlines of the particles from the manifold, we are led to interpret the lines of force of the electric field as a future-oriented time-like congruence with respect to the disformal metric. Roughly speaking, this means that disformal `time' starts at particles with positive charges and ends at particles with negative charges. Interestingly, working in the particular case of electric dipoles, we shall see that this interpretation gives rise to unexpected saddle-like singularities, where disformal time begins and ends simultaneously. Whether this can occur in GR for viable matter content remains (to the best of our knowledge) an open question.

The paper is divided as follows: in section II, we summarize the mathematical machinery behind disformal electrodynamics. In section III, we apply the mathematical setup to the case of electrostatic fields and discuss some geometrical objects of interest to the analysis of singularities. In section IV, we present the general ansatz for the disformal metric engendered by many charged particles along a line. In section V, we deal with the Newman-Penrose curvature invariants in some detail. In section VI, we restrict ourselves to the case of ``dipoles'' and show that distinct types of exotic singularities arise. We end up with the conclusions, pointing out the many possible developments that can be further explored.

\section{Mathematical tools}
Let $(\mathcal{M}, g_{ab})$ denote a four-dimensional Minkowski spacetime with metric signature convention $(+,-,-,-)$. Throughout, we shall be concerned with the electromagnetic field propagating in a source-free region $\mathcal{U}\subseteq\mathcal{M}$, such that
\begin{equation}\label{Maxwell}
F^{ab}{}_{;b}=0,\quad\quad\quad \star F^{ab}{}_{;b}=0.
\end{equation}
Here `` ; '' stands for covariant derivative compatible with $g_{ab}$ and
\begin{equation}
\star F^{ab}=\frac{1}{2}\varepsilon^{ab}{}_{cd}F^{cd}
\end{equation}
is the Hodge dual tensor with $\varepsilon_{0123}=\sqrt{-g}$. For the sake of concreteness, we assume all relevant fields to be sufficiently smooth on the region under consideration.

For any future-directed, timeline, and normalized field of observers, henceforth denoted by $X^{a}$, the electromagnetic tensor is decomposed as\footnote{Throughout, we adopt the following conventions: symmetrization is defined as $(ab)=ab+ba$ and anti-symmetrization as $[ab]=ab-ba$.}
\begin{eqnarray}
&&F^{ab}=E^{[a}X^{b]}+\varepsilon^{ab}{}_{cd}B^{c}X^{d},
\end{eqnarray}
with the electric and magnetic $3$-vectors given by
\begin{equation}
E^{a}=F^{ab}X_{b},\quad\quad\quad B^{a}=-\star F^{ab}X_{b}.
\end{equation}
Since these vectors are spacelike and automatically orthogonal to $X^{a}$, we shall write their squared norms as $E^{2}=-E^{a}E_{a}$ and $B^{2}=-B^{a}B_{a}$.

With the above conventions, the two independent field invariants read as
\begin{equation}
\psi\equiv\frac{1}{2}F_{ab}F^{ab}=B^{2}-E^{2},\quad\quad\quad\Upsilon\equiv\frac{1}{2}\star F_{ab}F^{ab}=EB\mbox{cos}\theta,
\end{equation}
with $\theta$ denoting the angle between the $3$-vectors. We define also the auxiliary invariant
\begin{equation}
\kappa\equiv\frac{1}{2}\sqrt{\psi^{2}+\Upsilon^{2}}=\frac{1}{2}\sqrt{E^{4}+B^{4}+2E^{2}B^{2}\mbox{cos}2\theta},
\end{equation}
which will play a prominent role in our discussion. An electromagnetic field is called \textit{regular} on $\mathcal{U}\subseteq\mathcal{M}$ if $\kappa\neq 0$ in this region and \textit{null} otherwise. From now on, we shall restrict ourselves solely to the regular case. For this case, it is well known that there exists a particular observer $X^{a}$ and a $3$-vector  $W^{a}$ such that
\begin{equation}
\fl\qquad
E^{a}=\alpha W^{a},\quad\quad\quad B^{a}=\beta W^{a},\quad\quad\quad W^{a}X_{a}=0,\quad\quad\quad W^{a}W_{a}=-1,
\end{equation}
for some $\alpha,\beta\in\mathbb{R}$ (see, for instance, the references \cite{Syn,Lan}). We shall call this particular observer a \textit{wrench observer} and, since $W^{a}$ is defined only up to a sign, we stick to the following convention: if $\psi\leq0$, we set $E^{a}$ and $W^a$ oriented at the same direction by choosing $\alpha>0$; if $\psi>0$, we set $B^{a}$ and $W^a$ oriented at the same direction by taking $\beta>0$. Clearly, the integral lines of $W^{a}$ define a smooth congruence of curves on the region under consideration. Henceforth, we call them simply by \textit{field lines}.

The energy-momentum tensor of the electromagnetic field is given, in appropriate units, by
\begin{equation}
T^{ab}=F^{a}{}_{c}F^{cb}+\frac{\psi}{2}g^{ab}.
\end{equation}
Besides its trace-free property, it also satisfies the so-called Ruse identity
\begin{equation}\label{Ruse}
T^{ac}T_{cb}=\kappa^{2}\delta^{a}{}_{b},
\end{equation}
which shows invertibility for regular fields. We leave it for the reader to verify that, since $\theta=0$ for a wrench observer, the energy-momentum tensor admits the following convenient representation \cite{Goulart21}
\begin{equation}\label{wrencht}
T_{ab}=\kappa[2(X_{a}X_{b}-W_{a}W_{b})-g_{ab}].
\end{equation}
From this equation, it is straightforward to obtain the algebraic relations
\begin{equation}
T_{ab}X^{b}=\kappa X_{a},\quad\quad\quad T_{ab}W^{b}=\kappa W_{a},
\end{equation}
showing that the wrench observer and its corresponding $3$-vector are actually eigenvectors of the energy-momentum tensor associated with the eigenvalue $\kappa$. Similarly, for any pair of linearly independent spacelike vectors $U^{a}$ and $V^{a}$, simultaneously orthogonal to $X^{a}$ and $W^{a}$, there follow
\begin{equation}
T_{ab}U^{b}=-\kappa U_{a},\quad\quad\quad T_{ab}V^{b}=-\kappa V_{a}.
\end{equation}
Hence, at each spacetime point, the energy-momentum tensor of a regular field defines what in the language of Schouten is called a \textit{two-bladed} structure \cite{Sch}. This induces a decomposition of the tangent space $T_{p}\mathcal{M}= A_{p}\oplus B_{p}$, with the 2-dimensional blades defined by
\begin{equation}
 A_{p}=\mbox{span}\{X,W\}_{p},\quad\quad\quad B_{p}=\mbox{span}\{U,V\}_{p}.
 \end{equation}

Now, suppose $\mathcal{U}\subseteq M$ is a source-free region where $F_{ab}$ is regular and consider the tensor
\begin{equation}\label{disformal}
\tilde{g}_{ab}\equiv-\kappa^{-1}T_{ab}
\end{equation}
Since $\tilde{g}_{ab}$ is non-degenerate by construction it is invertible and we may think of it as defining a new pseudo-Riemannian metric on the region. A rather curious result is the following: if $F_{ab}$ satisfies Maxwell equations on $(\mathcal{U},g_{ab})$ it also satisfies them on $(\mathcal{U},\tilde{g}_{ab})$. This is the disformal invariance of vacuum electrodynamics discussed in \cite{Goulart13,Goulart21}. Combining the equations (\ref{wrencht}) and (\ref{disformal}), we construct the \textit{disformal metric} and its corresponding inverse as
\begin{eqnarray}\label{covdisf}
&&\tilde{g}_{ab}=g_{ab}+2(W_{a}W_{b}-X_{a}X_{b}),\\\label{contdisf}
&&\tilde{g}^{ab}=g^{ab}+2(W^{a}W^{b}-X^{a}X^{b}),
\end{eqnarray}
from which we define the \textit{disformal line element}
\begin{equation}
d\tilde{s}^{2}=\tilde{g}_{ab}dx^{a}dx^{b}.
\end{equation}

\begin{figure}[ht]
    \centering
    \includegraphics[width=0.9\linewidth]{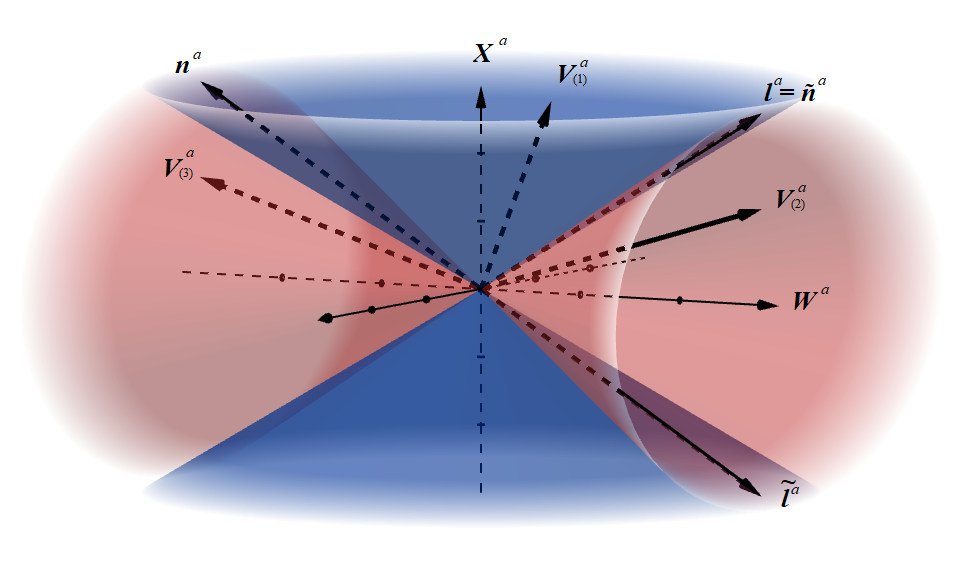}
    \caption{Light-cones for the base (blue) and disformal (red) metrics. Note that an arbitrary time-like vector $V_{(1)}^a$ with respect to $g_{ab}$ is a space-like one for $\tilde{g}_{ab}$, while future and past oriented time-like vectors for $\tilde{g}_{ab}$, given respectively by the representative vectors $V_{(2)}^a$ and $V_{(3)}^a$, are both space-like vectors for $g_{ab}$.}
    \label{fig_cones}
\end{figure}

Another useful way to write down the disformal metric is by introducing a tetrad basis of light-like vectors $\mathcal{B}=(l^a,n^a, m^a, \bar{m}^a)$ for the Minkowski spacetime and finding out the corresponding basis $\tilde{\mathcal{B}}$ associated with $\tilde{g}_{ab}$. To do so, we first define the vectors $l^a$ and $n^a$ in terms of $X^a$ and $W^a$ as
\begin{equation}
l^a=\frac{1}{\sqrt{2}}(X^a+W^a),\quad\mbox{and}\quad n^a=\frac{1}{\sqrt{2}}(X^a-W^a).\label{ln_X_W}
\end{equation}
Then, we choose a complex light-like vector $m^a$ mutually orthogonal to $l^a$ and $n^a$ and its complex conjugate $\bar{m}^a$ to fulfill the basis. In this representation, the Minkowski metric can be written as
\begin{equation}
\label{g_ab_null}
g_{ab}=l_a n_b +l_b n_a - m_a\bar{m}_b - m_b\bar{m}_a.
\end{equation}
Substituting this expression into the equation (\ref{covdisf}), with the definitions (\ref{ln_X_W}), we get the same decomposition for the disformal metric in terms of the basis $\tilde{\mathcal{B}}$, following the map \cite{Goulart21}
\begin{eqnarray}
\tilde{l}^a=-n^a,\quad \tilde{n}^a=l^a,\quad \mbox{and} \quad \tilde{m}^a=m^a,\label{map_contr}\\[1ex]
\tilde{l}_a=n_a,\quad \tilde{n}_a=-l_a,\quad \mbox{and} \quad \tilde{m}_a=m_a.\label{map_cov} \end{eqnarray}
Thus, this transformation can be interpreted as a point-dependent rotation of $\pi/2$ of the light-like vectors lying in the first blade, as illustrated by the figure (\ref{fig_cones}) (and a rotation of $-\pi/2$ of the light-like covectors). Contrarily to the case of disformal transformations involving a single vector \cite{Carvalho16}, the time orientation of the base metric is not preserved by the disformal transformation given by the equation (\ref{covdisf}). From the figure (\ref{fig_cones}), we see that time-like vectors with respect to $g_{ab}$ are space-like ones for $\tilde{g}_{ab}$.

Due to the above results, it is direct to show that the disformal transformation satisfies the following:
\begin{itemize}
\item{it is volume-preserving i.e., $\mbox{det}(g_{ab})=\mbox{det}(\tilde{g}_{ab})$};
\item{it exchanges the roles of space and time on the first blade i.e., $\tilde{\boldsymbol{g}}(W,W)=-\tilde{\boldsymbol{g}}(X,X)=1$};
\item{it does not affect the geometry of the second blade, since $\tilde{g}_{ab}|_{B}=g_{ab}|_{B}$};
\item{It satisfies a group structure, i.e., other disformal metrics can be constructed from a given one following a prescribed procedure \cite{Goulart13}.}
\end{itemize}

\section{Disformal curvature induced by electrostatic fields}

In this section we shall explore some general features of the curvature tensor associated to disformal metrics induced by electrostatic fields. To do so, we start by considering a wrench observer, such that
\begin{equation}\label{hypothesis}
X_{a;b}=0,\qquad F_{ab;c}X^{c}=0,\quad\mbox{and}\quad \star F_{ab;c}X^{c}=0.
\end{equation}
The first equation above means that the kinematics of $X^{a}$ is trivial, i.e., the vector field describes inertial observers without acceleration, expansion, shear and vorticity. The remaining equations imply that the electromagnetic field does not change along the world lines of $X^{a}$, as required by staticity. The electromagnetic field tensor then reads as
\begin{equation}
F^{ab}=EW^{[a}X^{b]},
\end{equation}
and will be regular as far as $E\neq 0$. The source-free Maxwell equations (\ref{Maxwell}), then give
\begin{equation}\label{maxdec}
W^{a}{}_{;a}=-C_{a}W^{a},\qquad W_{[a;b]}=C_{[a}W_{b]},\quad\mbox{and}\quad C_{a}\equiv \ln(E)_{;a}.
\end{equation}
From now on, we assume that $\mathcal{U} \subseteq \mathcal{M}$ is constructed by deleting from the spacetime the set consisting of the world lines of all charged particles generating the field plus the regions where the electric field possibly vanishes.

Introducing the Christoffel symbols as usual, we recall that for any pair of non-degenerate symmetric tensors $g_{ab}$ and $\tilde{g}_{ab}$, there follows the tensorial relation (see, for instance, the reference \cite{Wal})
\begin{equation}
C^{a}{}_{bc}\equiv\tilde{\Gamma}^{a}{}_{bc} - \Gamma^{a}{}_{bc} = \frac{1}{2}\tilde{g}^{ad}(\tilde{g}_{db;c}+\tilde{g}_{cd;b}-\tilde{g}_{bc;d}).
\end{equation}
By noticing that $g_{ab;c}=0$ and $\tilde{g}_{ab;c}X^{c}=0$, a straightforward calculation using equations (\ref{covdisf}) and (\ref{contdisf}) gives
\begin{equation}\label{ctensor}
C^{a}{}_{bc}=2\left\{r^{a}{}_{(b}W_{c)}-W^{a}d_{bc}-a^{a}W_{b}W_{c}\right\},
\end{equation}
where
\begin{equation}
r_{ab}\equiv\frac{1}{2}h_{a}{}^{c}h_{b}{}^{d}W_{[c;d]},\qquad d_{ab}\equiv\frac{1}{2}h_{a}^{c}h_{b}{}^{d}W_{(c;d)},\qquad a_{a}\equiv W_{a;b}W^{b},
\end{equation}
with the projector defined as
\begin{equation}
 h_{ab}\equiv g_{ab}-X_{a}X_{b}+W_{a}W_{b}.
 \end{equation}
 One easily verifies that
 \begin{equation}
  h_{ab}=h_{ba},\quad h_{ab}X^{b}=h_{ab}W^{b}=0,\quad h_{ac}h^{c}{}_{b}=h_{ab},\quad h^{a}{}_{a}=2.
  \end{equation}
By construction, one has the algebraic relations
\begin{eqnarray}\label{algprop}
&&r_{ab}W^{b}=0,\qquad d_{ab}W^{b}=0,\qquad a_{a}W^{a}=0,\\
&&r_{ab}X^{b}=0,\qquad d_{ab}X^{b}=0,\qquad a_{a}X^{a}=0,
\end{eqnarray}
which uniquely describe the kinematical parameters of the field lines spanned by $W^{a}$. Here $r_{ab}$ is the rotation, $d_{ab}$ is the distortion and $a_{a}$ is the acceleration. Furthermore, according to the above conventions, one has the trace relation $d\equiv d^{a}{}_{a}=W^{a}{}_{;a}$, which is nothing but the expansion of the spacelike congruence.

Now, since $W^{a}$ is constrained to satisfy the equations (\ref{maxdec}), the rotation tensor must vanish i.e., $r_{ab}=0$. We then obtain the simplified relation
\begin{equation}\label{lastc}
C^{a}{}_{bc}=-2\left\{W^{a}d_{bc}+a^{a}W_{b}W_{c}\right\}.
\end{equation}
 With this expression, we can seek for the condition of mutual geodesics in this context. A straightforward calculation shows that any affinely parametrized curve $x^{a}(\lambda)$ satisfying the geodesic equation in $g_{ab}$ will also be a geodesic in $\tilde g_{ab}$ whenever\\
 \begin{equation}
C^{a}{}_{bc}V^bV^c=f(x^d)V^a,\quad\quad \quad V^{a}=\frac{dx^a}{d\lambda},
\end{equation}
for some function $f(x^d)$ of the spacetime coordinates. In particular, from Eq. (\ref{lastc}), one sees that all time-like geodesics in $g_{ab}$, whose tangent vector is orthogonal to $W^a$, are also geodesics in $\tilde g_{ab}$, parameterized by the same parameter. A similar reasoning can be applied to the search for mutual Killing vectors, namely, a mutual Killing vector must satisfy
\begin{equation}
\label{mutual_killing}
C^{a}{}_{bc}\chi_a=0,
\end{equation}
where $\chi_a$ is a Killing covector with respect to the physical metric.

Finally, since we are dealing with a base Minkowski spacetime, it follows that the disformal curvature tensor reads as
\begin{equation}
\tilde{R}^{a}{}_{bcd}=C^{a}{}_{b[c;d]}+C^{a}{}_{e[d}C^{e}{}_{bc]}.
\end{equation}
Similarly, by noticing that $C^{a}{}_{ba}=0$, we obtain the disformal Ricci tensor and the scalar curvature as
\begin{equation}
\tilde{R}_{ab}=-C^{c}{}_{ab;c}+C^{c}{}_{db}C^{d}{}_{ac},\quad\mbox{and}\quad \tilde{R}=-\tilde{g}^{ab}C^{c}{}_{ab;c}.
\end{equation}
Now, using Eqs. (\ref{algprop}) and (\ref{lastc}), it is a direct calculation to show that
\begin{equation}
\label{disf_curv_scalar}
\tilde{R}=2(d'+d^{2}+a^{c}_{\phantom a;c}),
\end{equation}
with $d'\equiv d_{;a}W^{a}$. Recalling that $d=-C_{a}W^{a}$ due to the equation (\ref{maxdec}), the equation (\ref{disf_curv_scalar}) can be expressed in terms of the electric field and the wrench vector, as follows
\begin{equation}
    \tilde{R}=2\left\{-\left[(\ln E)_{; b} W^b\right]_{; a} W^a+\left[(\ln E)_{; a} W^a\right]^2+\left(W^a{}_{; b} W^b\right)_{; a}\right\},
\end{equation}
suggesting that there will be a singularity whenever one approaches a region of the spacetime where $E\rightarrow\infty$. As we shall see later, the limit $E\rightarrow 0$ can also lead to a singularity, if some of its derivatives do not vanish faster than $E$ itself. This will be precisely the case of a saddle-like singularity, as we mentioned before. Otherwise, the disformal metric will approach the Minkowski one when $E\rightarrow 0$.

\section{Particles along a line}
Let $x^{\mu}=(t,\rho,\varphi, z)$ describe the usual cylindrical coordinates, for which the infinitesimal line element of Minkowski spacetime reads as
\begin{equation}
\label{cyl_metric}
ds^2=dt^2-d\rho^2-\rho^{2}d\varphi^2-dz^2.
\end{equation}
Consider $n\in\mathbb{N}$ static point particles with electric charges $q_{1},q_{2},...,q_{n}\in\mathbb{R}$, located along the $z$-axis. If their positions are represented generically by $p_{1},p_{2},...,p_{n}\in\mathbb{R}$, the corresponding four-potential is given by $A_{\mu}(\rho,z)=(\phi(\rho,z),0,0,0)$, where
\begin{equation}\label{pot1}
\phi(\rho,z)=\sum_{k=1}^n \phi_{k}(\rho,z)\label{potential}
\end{equation}
with
\begin{equation}\label{pot2}
\phi_{k}=q_{k}/\zeta_{k},\quad\quad\quad \zeta_{k}^{2}\equiv \rho^{2}+(z-p_{k})^{2},\quad\quad\quad k=1,2,...,n\label{zeta}.
\end{equation}
For the sake of compactness, we write the corresponding Faraday tensor in the language of differential forms as
\begin{equation}
\boldsymbol{F}=\frac{1}{2}F_{ab}dx^{a}\wedge dx^{b}=\partial_{\rho}\phi\ d\rho\wedge dt+\partial_{z}\phi\ dz\wedge dt\label{2formelec},
\end{equation}
and its Hodge dual as
\begin{equation}
\star \textbf{F}=\frac{1}{2}\star F_{ab}dx^{a}\wedge dx^{b} = \partial_{z}\phi\ d\varphi\wedge dz+\partial_{\rho}\phi\ d\rho\wedge d\varphi\label{2formelecdual}
\end{equation}
with
\begin{equation}
\label{part_phi}
\partial_{\rho}\phi=-\rho\sum_{k=1}^n q_{k}\zeta_{k}^{-3},\quad\quad\quad \partial_{z}\phi=-\sum_{k=1}^n q_{k}\zeta_{k}^{-3}(z-p_{k}).
\end{equation}
The corresponding wrench $3$-vector $W^a$ associated to the trivial wrench observer $X^a=\delta^a_{t}$ is given by
\begin{equation}
W^a=-\frac{1}{|\nabla\phi|}(a\,\delta^a_{\rho}+b\,\delta^a_{z})
\label{W},
\end{equation}
with $a\equiv \partial_{\rho}\phi$, $b\equiv \partial_{z}\phi$ and $|\nabla\phi|=\sqrt{a^2+b^2}$, for conciseness. We then construct our region of interest $\mathcal{U}\subseteq\mathcal{M}$ by excising from the manifold the world lines of the charged particles and all regions where $a^{2}+b^{2}$ possibly vanishes.

Using Eqs (\ref{covdisf}) and (\ref{cyl_metric}), one obtains the disformal infinitesimal line element as
\begin{eqnarray}\label{disfmet}
d\tilde{s}^{2}&=&\frac{1}{|\nabla\phi|^2}[(ad\rho+bdz)^{2}-(b d\rho-adz)^{2}]-dt^{2}-\rho^{2}d\varphi^{2}.
\end{eqnarray}
We now seek a coordinate system where the metric above is diagonal. The first candidate as a new ``time'' coordinate is the potential $\phi(\rho,z)$ itself. We notice that its level sets (the equipotential hypersurfaces) are generated by rotating a family of plane curves about the z-axis and foliates the manifold with spacelike codimension one surfaces. The other coordinate is obtained by defining the following non-exact 1-form
\begin{equation}
\boldsymbol{\omega}\equiv b d\rho-adz.
\end{equation}
By taking its exterior derivative, one gets
\begin{equation}
d\boldsymbol{\omega}=-(\partial^{2}_{\rho}\phi+\partial^{2}_{z}\phi)d\rho\wedge dz=(\rho^{-1}\partial_{\rho}\phi) d\rho\wedge dz,
\end{equation}
where we have used the Laplace equation in cylindrical coordinates. From this equation, one easily sees that the coordinate $\rho$ is an integrating factor for $\boldsymbol{\omega}$, i.e., $d(\rho\boldsymbol{\omega})=0$. In other words, there exists a function $\xi$ such that
\begin{equation}
\label{def_part_xi}
d\xi=\rho\boldsymbol{\omega},\quad\rightarrow\quad\partial_{\rho}\xi=\rho \partial_{z}\phi,\quad \partial_{z}\xi=-\rho \partial_{\rho}\phi.
\end{equation}
Remarkably, it admits the simple explicit solution
\begin{equation}\label{W1}
\xi=\sum_{k=1}^n \xi_{k},\quad\quad\quad \xi_{k}=(z-p_{k})\phi_{k}\label{generalxi}.
\end{equation}

The above considerations  suggest that we stick to a new coordinate system $\tilde{x}^{\mu}=(t,\phi,\varphi,\xi)$, replacing the old coordinates $\rho$ and $z$. For the system $\tilde{x}^{\mu}$ to be well-defined, we must impose
\begin{equation}
\mbox{det}\left[\begin{array}{cc}
	\partial_{\rho}\phi & \partial_{z}\phi  \\
	\partial_{\rho}\xi & \partial_{z}\xi \\
	\end{array}\right]=-\rho|\nabla \phi|^{2}\neq 0.
\end{equation}
Henceforth, we shall call the above coordinate system by \textit{equipotential coordinates}. The main advantage of using it relies upon the following relation
\begin{equation}
d\rho^2 + dz^2=\frac{1}{|\nabla \phi|^{2}}[d\phi^2+\rho^{-2}d\xi^2],
\end{equation}
which may be easily verified using the equations (\ref{part_phi}) and (\ref{def_part_xi}). Taking this into account in the equation (\ref {cyl_metric}), gives
\begin{equation}
ds^2=dt^2-\frac{1}{|\nabla \phi|^{2}}[d\phi^2+\rho^{-2}d\xi^2]-\rho^{2}d\varphi^2,
\end{equation}
where $\rho$ is to be understood as an implicit function of $\phi$ and $\xi$. In its turn, the disformal line element (\ref{disfmet}) written in the equipotential coordinates becomes
\begin{equation}
\label{disfmet_eq_pot}
d\tilde{s}^2=\frac{1}{|\nabla \phi|^{2}}[d\phi^2-\rho^{-2}d\xi^2] -dt^2 -\rho^{2}d\varphi^2.
\end{equation}
Note that the translations with respect to the $t$-coordinate and the rotations around the $\varphi$-coordinate are clearly symmetries preserved by this disformal metric, which can also be verified by the equation (\ref{mutual_killing}). Furthermore, due to the diagonal form of the disformal metric in this coordinate system, the calculation of the scalar curvature given by Eq.\ (\ref{disf_curv_scalar}) is rather simplified, yielding
\begin{equation}
\fl\qquad
\tilde{R}=2\left\{(\partial_{\phi}E)^2 - E\partial_{\phi\phi}E + \rho^2[E\partial_{\xi\xi}E - (\partial_{\xi}E)^2] + 2\rho E \partial_{\xi}\rho\partial_{\xi}E\right\}.
\end{equation}
In fact, as one approaches a particular charge of the configuration, the electric field tends to represent a monopole, and thus, its derivatives with respect to $\xi$ become negligible (the same occurs far away from the charges). Therefore, the existence of a singularity is basically driven by $E$ and its derivatives with respect to $\phi$. In other words, there will be a singularity in the scalar curvature at all ``charge locations'' because the derivatives of the electric field diverge there. Similarly, $\tilde R$ goes to zero far away from the sources, since $E$ and its derivatives vanish in this regime. However, it is easy to realize that as one approaches infinity in physical space-like directions, asymptotically rectangular coordinates do not exist\footnote{see \cite{misner1967taub} for a similar situation in the context of TAUB-NUT spaces.}. Such qualitative analysis will be confirmed below when we apply this framework to the case of two point charges.

\section{Newman-Penrose curvature invariants}
The NP formalism is a special case of the tetrad formalism, where the vector basis is constructed with four light-like vectors (two real, and a complex-conjugate pair). Usually, the curvature invariants in the NP formalism provide information about the profile of the gravitational radiation in the asymptotic regime, as a result of the peeling-off Theorem. This formalism is also helpful in finding symmetries of the spacetime or in the determination of whether the metric is algebraic special, in the language of the Petrov classification.

In the case of $n$ point charges displaced along the z-axis, the curvature invariants can be calculated using the Newman-Penrose (NP) formalism, where the vectors of the null tetrad basis $\tilde{\mathcal{B}}$ can be explicitly obtained from equations (\ref{ln_X_W})-(\ref{map_cov}) and (\ref{W}). That is,
\begin{eqnarray}
\fl\qquad
\tilde{l}^a=\frac{1}{\sqrt{2}}\left(1,\frac{a}{|\nabla \phi|},0,\frac{b}{|\nabla \phi|}\right),\qquad\tilde{n}^a=\frac{1}{\sqrt{2}}\left(1,-\frac{a}{|\nabla \phi|},0,-\frac{b}{|\nabla \phi|}\right),\\[1ex]
\fl\qquad
\tilde{m}^a=\frac{1}{\sqrt{2}}\left(0,\frac{b}{|\nabla \phi|},-\frac{i}{\rho},-\frac{a}{|\nabla \phi|}\right),\qquad \bar{\tilde{m}}^a=\frac{1}{\sqrt{2}}\left(0,\frac{b}{|\nabla \phi|},\frac{i}{\rho},-\frac{a}{|\nabla \phi|}\right).
\end{eqnarray}\\
Computing the spin coefficients of the disformal metric with the help of computational algebra, we get the following non-trivial ones
\begin{eqnarray}
\fl\qquad&&\tilde{\pi}=\tilde{\nu}=-\tilde{\tau}=-\tilde{\kappa}=\frac{\sqrt{2}\,a^2}{4\,|\nabla \phi|^{3}}\left(a\,\partial_{\rho} + b\, \partial_{z}\right)\left(\frac{b}{a}\right),\qquad\tilde{\beta}=-\tilde{\alpha}=\frac{\sqrt{2}\,b}{4\,\rho\,|\nabla \phi|},\nonumber\\[1ex]
\fl\qquad&&\tilde{\lambda}=-\tilde{\sigma}=\tilde{\mu}+2\frac{a}{b}\tilde{\beta}=-\tilde{\varrho}+2\frac{a}{b}\tilde{\beta}=\frac{\sqrt{2}\,a^2}{4\,|\nabla \phi|^{3}}\left(a\partial_{z} - b\partial_{\rho}\right)\left(\frac{b}{a}\right)+\frac{\sqrt{2}\,a}{4\,\rho\,|\nabla \phi|}.
\end{eqnarray}

From these coefficients, we calculate the corresponding Ricci and Weyl NP invariants of the disformal metric. Yielding
\begin{eqnarray}
\fl\qquad&&
\frac{\tilde{R}}{4}=\frac{ a^{2}b(\partial_{zz}-\partial_{\rho\rho})b}{|\nabla\phi|^{4}} + \frac{a(a^{2} - b^{2}) \partial_{z\rho}b}{|\nabla\phi|^{4}} - \frac{ab^{2}(\partial_{zz}-\partial_{\rho\rho})a}{ |\nabla\phi|^{4}}-\frac{b(a^{2}- b^{2})\partial_{z\rho}a}{|\nabla\phi|^{4}}\nonumber\\[1ex]
\fl\qquad&&
+ 2\frac {ab(a^{2} - 3b^{2}) (\partial_{z}a)\, \partial_{\rho}a}{ |\nabla\phi|^{6}} -2\frac {ab \left( 3\, a^{2}- b^{2} \right)  \left( \partial_{z}b\right) \partial_{\rho}b}{ |\nabla\phi|^{6}} \nonumber\\[1ex]
\fl\qquad&&
+4{\frac {ab( a^{2}- b^{2})[(\partial_{\rho}b)\partial_{\rho}a+(\partial_{z}b)\partial_{z}a]}{ |\nabla\phi|^{6}}} - \frac { \left(  a^{4}-6\, a^{2} b^{2}+ b^{4} \right) [(\partial_{\rho}b)\partial_{z}a+(\partial_{\rho}a)\partial_{z}b]}{ |\nabla\phi|^{6}} \nonumber\\[1ex]
\fl\qquad&&
+{\frac { b^{2} \left( 3\, a^{2}- b^{2} \right)  [( \partial_{z}a)^{2}-( \partial_{\rho}a)^{2}]}{ |\nabla\phi|^{6}}}+ \frac{a^{2} (a^{2}-3\, b^{2})  [( \partial_{z}b)^{2}-(\partial_{\rho}b)^2]}{ |\nabla\phi|^{6}}\nonumber\\[1ex]
\fl\qquad&&
+2{\frac {ab(b\partial_{\rho}a-a\partial_{\rho}b)}{\rho\, |\nabla\phi|^{4}}} +{\frac {(a^{2} - b^{2}) (a\partial_{z}b-b\partial_{z}a)}{ \rho\,|\nabla\phi|^{4}}}\\[1ex]
\fl\qquad&&
\tilde{\Phi}_{11}=\frac{a^3}{2\,\rho |\nabla\phi|^{4}}( a\partial_{z}-b\partial_{\rho})\left(\frac{b}{a}\right),\\[1ex]
\fl\qquad&&
\tilde{\Psi}_{0}=\tilde{\Phi}_{11}-\frac{\tilde{R}}{8}-\frac{a^2\,b}{\rho|\nabla\phi|^4}(a\partial_{\rho} + b \partial_{z})\left(\frac{b}{a}\right),\\[1ex]
\fl\qquad&&
\tilde{\Psi}_{2}=-\frac{\tilde{\Psi}_{0}}{3}-\frac{a^2}{\rho|\nabla\phi|^2}\partial_{z}\left(\frac{b}{a}\right),
\end{eqnarray}
The other non-zero invariants are obtained through the relations $\tilde{\Psi}_4=-\tilde{\Phi}_{02}=\tilde{\Psi}_0$ and $\tilde{\Phi}_{00}=\tilde{\Phi}_{22}=\tilde{\Phi}_{11}-\tilde{R}/8$, applicable for any potential given by the equation (\ref{pot1}). It should also be remarked that the NP invariants above could be calculated from the spin coefficients of the background metric according to the formulas derived in the reference \cite{Goulart21}.

\section{The case of two point charges}
Now, we return to the discussion of the equipotential coordinates, under the restriction of having only two charged particles. To find explicitly the cylindrical coordinates $\rho$ and $z$ in terms of the equipotential ones $\phi$ and $\xi$, we recall the auxiliary functions
$\zeta_{1,2}^2(\rho,z)=\rho^2+(z-p_{1,2})^2$, and, without loss of generality, set the charges symmetrically displaced with respect to the plane $z=0$, such that $p_1=-p_2\equiv p$. This can be seen as a coordinate transformation and the inverse relation is given by
\begin{equation}
\fl\qquad z(\zeta_1,\zeta_2)=\frac{1}{4p}(\zeta_2^2-\zeta_1^2),\quad\mbox{and}\quad \rho(\zeta_1,\zeta_2)=\sqrt{\frac{\zeta_2^2+\zeta_1^2}{2}-\frac{(\zeta_2^2-\zeta_1^2)^2}{16p^2}-p^2}\label{rho}.
\end{equation}
On the other hand, the equipotential coordinates can be written in terms of $\zeta_{1,2}$, as follows
\begin{equation}
\fl\qquad
\phi(\zeta_1,\zeta_2)=\frac{q_1}{\zeta_1}+\frac{q_2}{\zeta_2},\quad\mbox{and}\quad\xi=q_1\,\frac{z(\zeta_1,\zeta_2)-p}{\zeta_1}+q_2\,\frac{z(\zeta_1,\zeta_2)+p}{\zeta_2}.
\end{equation}
The attempt to find the coordinate transformation $\rho(\phi,\xi)$ and $z(\phi,\xi)$ requires the solution of the following quintic function for $\zeta_2$:
\begin{eqnarray}
\label{eq_zeta2}
\fl\qquad\frac{\phi^3}{q_2^3}\,\zeta_2^{5}-2\,\frac{\phi^2}{q_2^2}\,\zeta_2^{4} -\frac{\phi}{q_2}\left(4\,p\frac{\phi^2}{q_2^2} + 4\,\frac{\phi}{q_2}\frac{\xi}{q_2} + \frac{q_1^2}{q_2^2} - 1\right)\zeta_2^{3}+ 8\frac{\phi}{q_2}\left(2\,p\frac{\phi}{q_2} +\frac{\xi}{q_2}\right) \zeta_2^{2}\nonumber\\[1ex]
\fl\qquad-4\left( 5p\frac{\phi}{q_2}+ \frac{\xi}{q_2}\right) \zeta_2 + 8\,p=0,
\end{eqnarray}
together with $\zeta_1=\frac{q_1\zeta_2}{q_2+\phi \zeta_2}$. As far as we have tried, the techniques to encounter the roots of this polynomial equation seem hopeless. Therefore, in order to proceed with the analysis some restriction in the charge configuration is required, and it shall be done in the next steps.

\subsection{The electric dipole}
\begin{figure}[ht]
\includegraphics[scale=0.3]{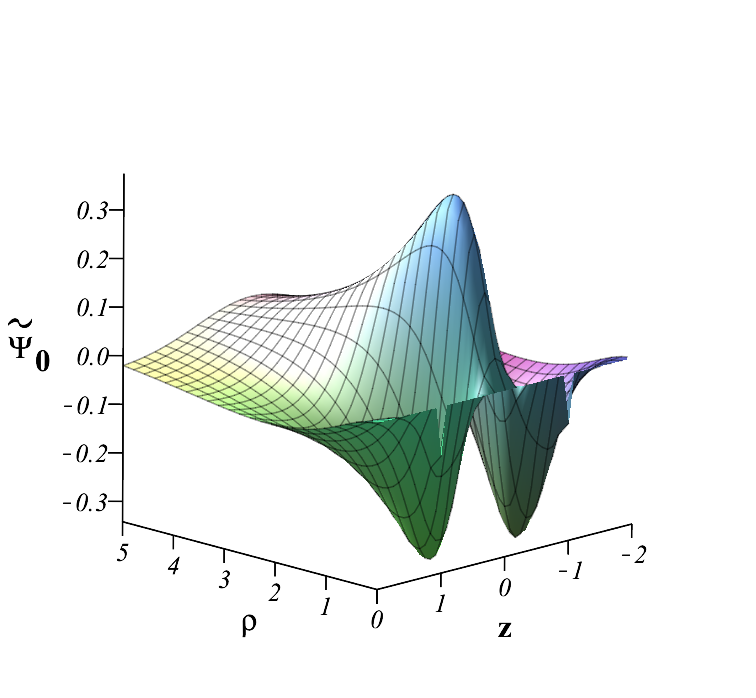}
\includegraphics[scale=0.3]{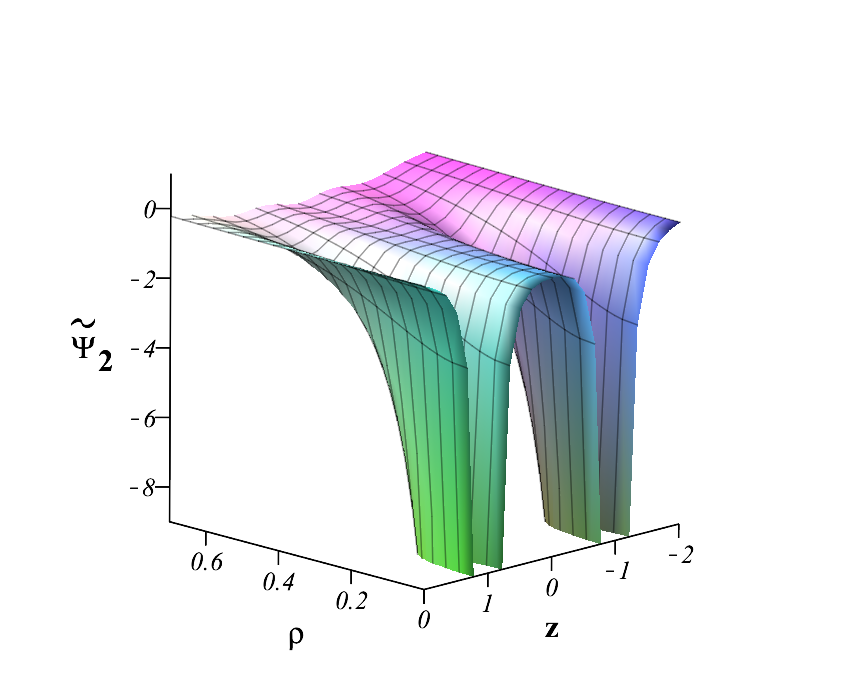}
\includegraphics[scale=0.3]{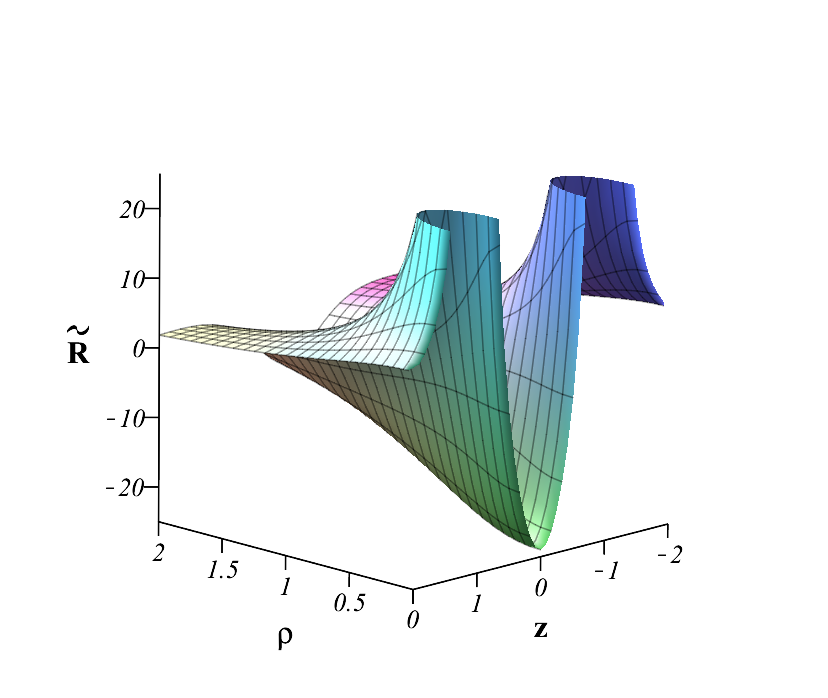}
\hspace{.5cm} \includegraphics[scale=0.3]{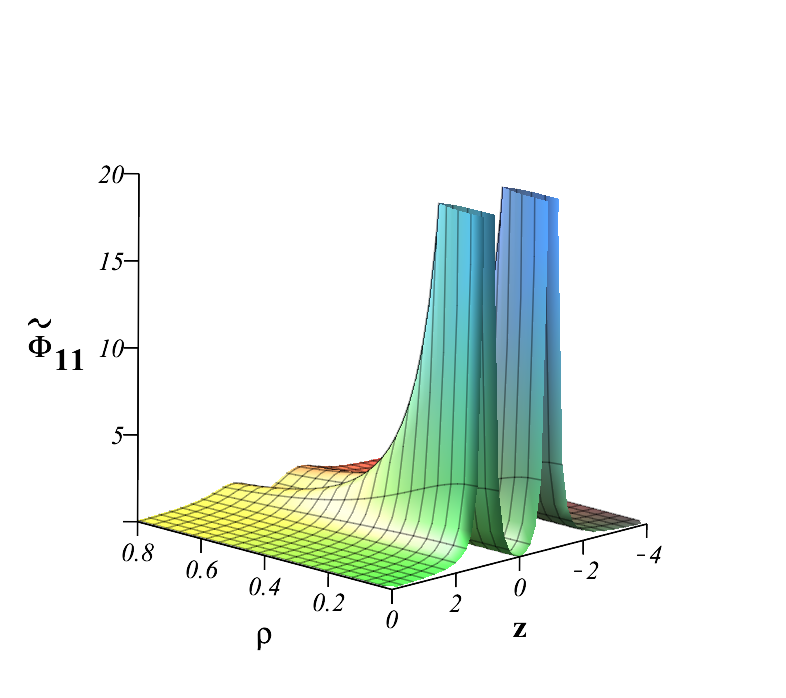}
\caption{Ricci and Weyl invariants for the electric dipole case. The choice of the parameters are $q_1=-q_2=1$ and $p=1$. For large $\rho$ and $z$, it is evident the asymptotically flat behavior of the invariants. $\tilde{\Psi}_2$, $\tilde{R}$ and $\tilde{\Phi}_{11}$ show the singularities in the curvature at the charge positions, while $\tilde{\Psi}_0$ is regular everywhere.}
\label{figdipole}
\end{figure}

The case of an electric dipole corresponds in setting $q_1=-q_2\equiv q$ in the equations above. It does not simplify substantially equation (\ref{eq_zeta2}) to the point of being solvable, but it gives interesting behaviors for the NP invariants as depicted in figure (\ref{figdipole}). Despite $\tilde\Psi_0$ being regular everywhere, all the other invariants diverge quadratically at the positions of the charges. $\tilde{\Psi}_2$ is non-positive while $\tilde{\Phi}_{11}$ is non-negative. In the vicinity of the singularities, their behaviors are given below, where we first expanded the invariants around the charges and, further, each coefficient of the series around $\rho=0$. Thus,\\
\begin{eqnarray}
\fl\qquad\tilde{R}^{(\pm)}&=&\frac{4}{\rho^{2}}+\frac{15}{2}\,\rho - \frac{25\,{\rho}^{2}}{4} \pm \left( \frac{8}{\rho} - 27\,\rho \right)( z \mp 1) - \left( \frac{4}{\rho^{4}}+
\frac {75}{4\,\rho} \right)( z \mp 1 )^{2}\nonumber\\[1ex]
\fl\qquad&&+ \mathcal{O}(( z \mp 1)^3),\\[1ex]
\fl\qquad\tilde{\Phi}_{11}^{(\pm)}&=&\frac{1}{2 \rho^{2}}+\frac{3 \rho}{16}-\frac{\rho^{2}}{8}\pm\frac{z\mp 1}{4 \rho}+\left(-\frac{1}{2 \rho^{4}}-\frac{15}{32 \rho}\right) \left(z\mp 1\right)^{2}+\mathcal{O}(( z \mp 1)^3),\\[1ex]
\fl\qquad\tilde{\Psi}_{0}^{(\pm)}&=&-\frac{3 \rho}{4}+\frac{15 \rho^{2}}{32}\pm \frac{75 \rho  \left(z\mp 1\right)}{32}+\frac{3 \left(z\mp 1\right)^{2}}{8 \rho}+\mathcal{O}(( z \mp 1)^3),\\[1ex]
\fl\qquad\tilde{\Psi}_{2}^{(\pm)}&=&-\frac{1}{3 \rho^{2}}+\frac{\rho}{8}-\frac{13 \rho^{2}}{96}\pm \left(\frac{1}{12 \rho}-\frac{9 \rho}{16}\right) \left(z\mp 1\right) +\left(\frac{1}{3 \rho^{4}}-\frac{5}{16 \rho}\right) \left(z\mp 1\right)^{2}\nonumber\\[1ex]
\fl\qquad&&+\mathcal{O}(( z \mp 1)^3),
\end{eqnarray}
where the superscript $+$ holds for the expansion around $z=+1$ and the superscript $-$ corresponds to the expansion around $z=-1$, (where we set $p=1$ for the sake of simplicity). It is easy to see that all invariants are asymptotically flat for large $\rho$ and $z$, a consequence of the local charge distribution used in the disformal transformation.

From Eq. (\ref{disfmet}) one sees that, according to the sign of the function $a^2-b^2$, the coordinates $\rho$ and $z$ exchange the role of a time coordinate for the disformal metric. A preliminary analysis indicated that the causal structure is very complicated in such coordinates and further investigation is left for future work. In terms of the equipotential coordinates, the interpretation is simpler, since $\phi$ and $\xi$ have a clear role in the metric as time and space coordinates, respectively. Another interesting point to notice is that the singularities are sorts of localized ``big-bangs'' and ``big-crunchs'', since the field lines work as time-like curves by definition. In other words, since the positive charges work as sources of the field lines, they may be interpreted as the origins of time, while the negative charges working as sinks for the field lines would be the ends of time. That is the reason for the analogy with the cosmological terms big-bang and big-crunch. Therefore, the incompleteness of causal curves, predicted by the singularity theorems, is a trivial consequence of disformal electrodynamics. Last but not least, it should also be remarked that the electric potential is not static anymore due to the change of roles of the cylindrical coordinates in the first blade, changing drastically the interpretation of the electromagnetic configuration under analysis without modifications in its functional form.

\subsection{Two identical charges}
\begin{figure}[!ht]
\includegraphics[scale=0.3]{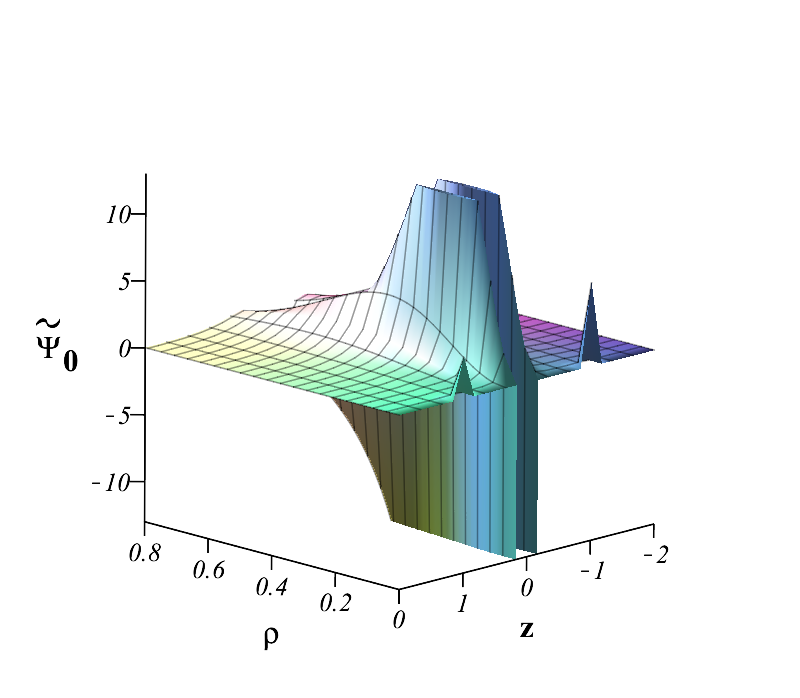}
\includegraphics[scale=0.3]{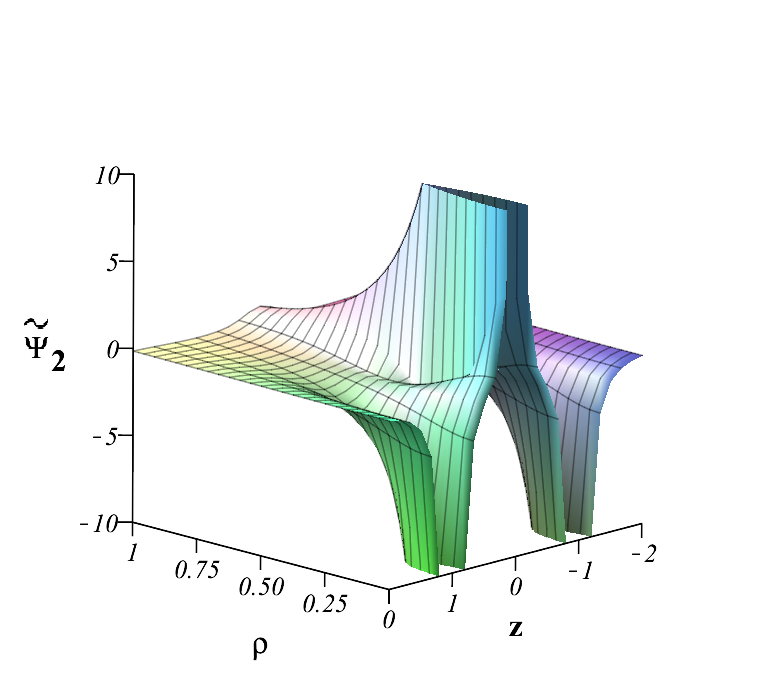}
\includegraphics[scale=0.3]{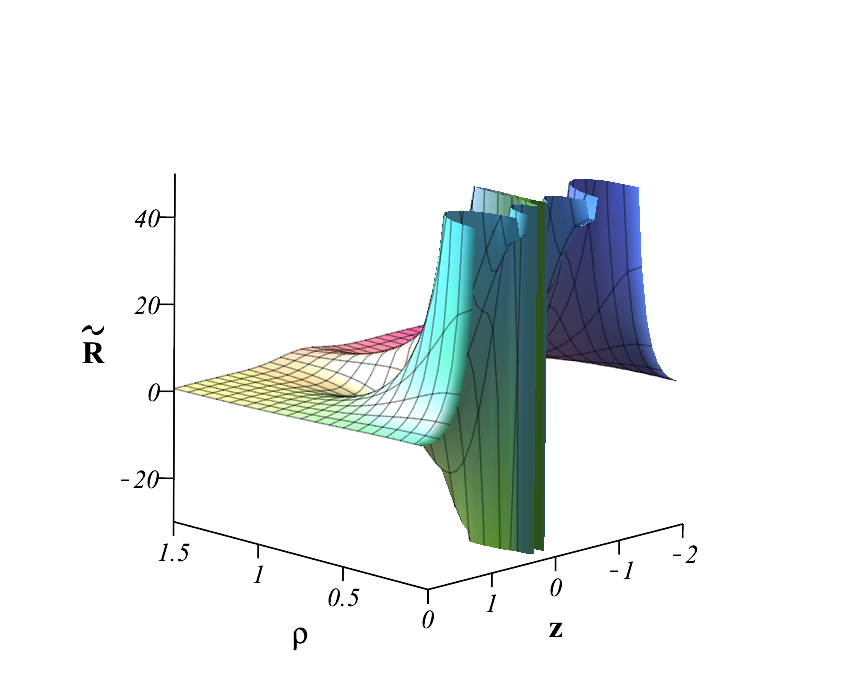}
\includegraphics[scale=0.3]{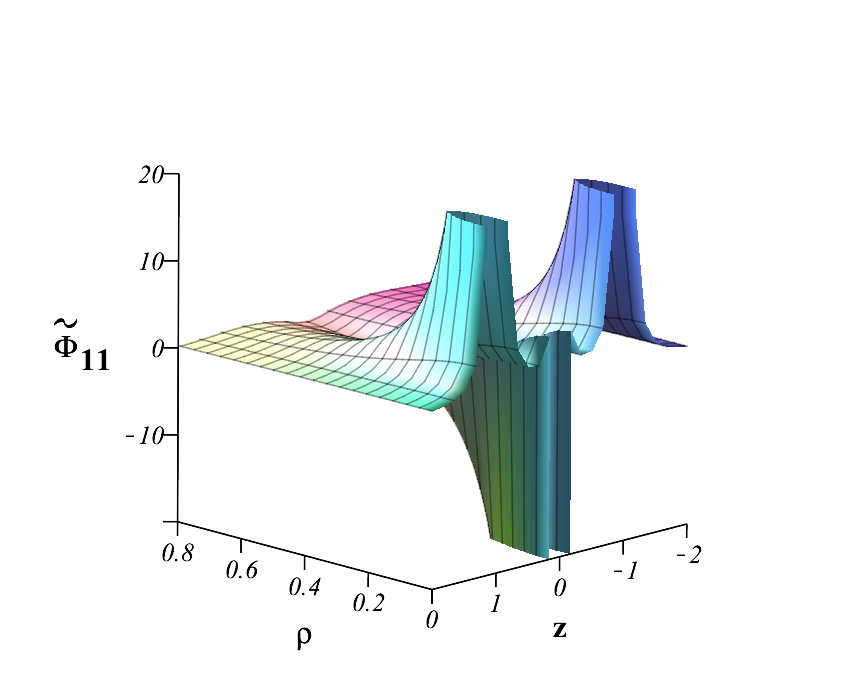}
\caption{Ricci and Weyl invariants for the case of the identical charges. The choice of the parameters is $q_1=q_2=1$ and $p=1$. Again, the invariants are all asymptotically flat in $\rho$ and $z$. All invariants contain at least one singularity either at the charge positions or at the origin.}
\label{figeqcharge}
\end{figure}

By considering the previous analysis, the calculations for the case of identical charges are straightforward, but the results are qualitatively distinct. From the figure (\ref{figeqcharge}), one sees the appearance of an extra singularity at the origin, such that the NP invariants around this point are
\begin{eqnarray}
\fl\qquad&&
\tilde{R}^{(0)}=\frac{16}{\rho^{2}}+\left(-\frac{592}{\rho^{4}}+\frac{1536}{\rho^{2}}\right) z^{2}+\mathcal{O}(z^3),\\
\fl\qquad&&
\tilde{\Phi}_{11}^{(0)}=-\frac{1}{\rho^{2}}+\left(\frac{10}{\rho^{4}}-\frac{24}{\rho^{2}}\right) z^{2}+\mathcal{O}(z^3),\\
\fl\qquad&&
\tilde{\Psi}_{0}^{(0)}=-\frac{3}{\rho^2} + \left(\frac{96}{\rho^4} - \frac{252}{\rho^2}\right)z^2+\mathcal{O}(z^3),\\
\fl\qquad&&
\tilde{\Psi}_{2}^{(0)}=\frac{5}{3 \rho^{2}}+\left(-\frac{104}{3 \rho^{4}}+\frac{88}{\rho^{2}}\right) z^{2}+\mathcal{O}(z^3).
\label{riccicharge}
\end{eqnarray}
Note that all of them have a quadratic divergence at the origin with respect to the $\rho$-coordinate. At the charges, the qualitative behavior of the invariants is the same as before, for instance, the Ricci scalar is
\begin{equation}
\fl
\tilde{R}^{(\pm)}=\frac{4}{\rho^{2}}-\frac{15 \rho}{2}-\frac{25 \rho^{2}}{4}\mp\left(\frac{8}{\rho}-27 \rho \right) \left(z\mp 1\right)
-\left(\frac{4}{\rho^{4}}-\frac{75}{4 \rho}\right) \left(z\mp 1\right)^{2}+\mathcal{O}(( z \mp 1)^3).
\end{equation}
Now the interpretation of the singularities is even more elaborate because the origin is a saddle point for the field lines. Therefore, along the z-axis one encounters three types of singularities: either two ``big-bangs'' ($q>0$) or two ``big-crunches'' ($q<0$), besides a saddle-like singularity at the origin where we can interpret the disformal time beginning and ending ``simultaneously'', because there are some field lines outgoing and others incoming depending on the direction one approaches the singularity. Along the $\rho$ axis the nature of the singularity at the origin will be the same as along the charges.

\subsection{Perfect dipole}
\begin{figure}[!ht]
\centering
\includegraphics[scale=0.5]{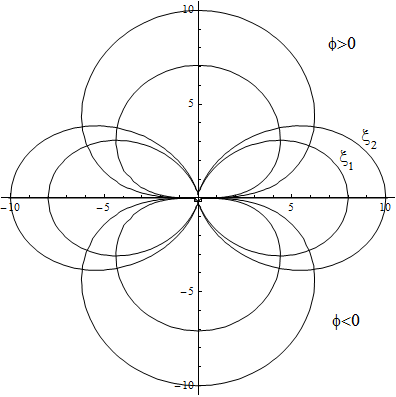}
\caption{Illustration of the dipole coordinates. We use the coordinates $x$ and $z$, setting $y=0$ and $M=2$ in the equation (\ref{phiperfectdi}) written in the Cartesian coordinates. Of course, the figure is symmetric with respect to rotations around the z-axis.}
\label{fig_dip_coord}
\end{figure}

A particularly interesting case is when the distances involved are much larger than the separation distance of the charges. Mathematically, it means $p\ll z$ with the final configuration representing the so-called perfect dipole. In this regime, we can expand the expressions for $\phi$ and $\xi$ for small $p$, obtaining
\begin{equation}
\label{phiperfectdi}
\phi(r,\theta)=\frac{M\,\cos\theta}{r^2},
\quad \mbox{and}\quad\xi(r,\theta)=-\frac{M\,\sin^2\theta}{r},
\end{equation}
which we have conveniently written in spherical coordinates, with $M=2pq$ as the dipole moment. According to the reference \cite{Carr1965}, it is possible now to invert the transformation, finding $r(\phi,\xi)$ and $\theta(\phi, \xi)$, because the expression corresponding to the equation (\ref{eq_zeta2}) reduces to a quartic polynomial. By making the identification $\beta\equiv\phi/M$ and $\alpha\equiv-M/\xi$, the inverse transformation is given by the equations (41) and (59) of the reference \cite{Carr1965}. The time plus space foliation of the manifold is depicted in figure \ref{fig_dip_coord}. The semi-plane $z>0$ corresponds to $\phi$ positive, while $z<0$ corresponds to $\phi$ negative. The coordinate dictated by the field lines is such that $\xi_1<\xi_2<0$.

The non-trivial NP curvature invariants simplify tremendously, and their explicit expressions are:
\begin{eqnarray}
    && \tilde{R}=\frac{12(27\cos^6\theta + 35 \cos^4\theta + 21 \cos^2\theta - 3)}{r^{2} (3\cos^2\theta +1 )^3},\\[1ex]
    && \tilde{\Phi}_{11}=\frac{9( \cos^2\theta + 1) \cos^2\theta}{r^{2}(3\cos^2\theta +1 )^2},\\[1ex]
    && \tilde{\Psi}_0= \frac{3(9\cos^6\theta + 7\cos^4\theta - 17 \cos^2\theta + 1)}{2\,r^{2} (3\cos^2\theta +1 )^3},\\[1ex]
    && \tilde{\Psi}_2=-\frac{27\cos^6\theta  + 37\cos^4\theta - 3\cos^2\theta + 3}{2\,r^{2} (3\cos^2\theta +1 )^3}.
\end{eqnarray}
The appealing polynomial structure of the invariants in terms of trigonometric functions, emphasizes the peculiar characteristic of the singularity in this case: according to the sign of the numerator, the invariants will either diverge to $+\infty$, $-\infty$ or vanish abruptly. In other words, from the analysis of the scalars, it is found that the disformal metric exhibits directional type singularities i.e., the singular behavior of the scalars depends upon the direction by which the singular point is approached. This is a rare behavior in solutions of general relativity \cite{gautreau1967directional, taylor2005unravelling}, but seems to be very common for disformal electromagnetic metrics: a consequence of the possible saddle character of field lines and the fact that these lines are, roughly speaking, time-like curves in the disformal metric.

\section{Concluding remarks}

We have taken a significant step towards understanding the nature of singularities that arise in disformal electrodynamics. Our study reveals that a point charge of electromagnetism in Minkowski spacetime generates a curvature singularity in the disformal metric. However, the behavior of the spacetime is far from simple. In fact, it is highly exotic compared to that of general relativity, especially concerning its causal structure and topology. In particular, for the cases we analyzed, the existence of singularities is immediately predictable because the straight field lines that connect the point charges are incomplete geodesics of the disformal metric.

Interestingly, when we describe the disformal metric in terms of the initial coordinate system of the base metric, we observe a peculiar behavior. Specifically, there are regions where the cylindrical coordinates $\rho$ and $z$ can exchange the role as time coordinate without resulting in divergences in the components of the disformal metric. This phenomenon may be a hint for the presence of closed time-like curves, but we deferred further investigation for future work.

Moreover, we expanded the NP invariants in a power series close to the singularities for the three cases of interest: the two point charges configuration with opposite signs, equal signs, and the perfect dipole. We observed a quadratic divergence of almost all NP invariants with respect to the coordinate $\rho$ in all cases, but its sign is not uniquely determined. Depending on the direction one takes the limit of each invariant, it goes to plus or minus infinity, or even zero. We suggested classifying the singularities as ``big-bangs'' (where field lines depart) or ``big-crunches'' (where field lines arrive), but the sign of the curvature stays undetermined. In this sense, our investigation supports the idea of the singularity theorems, where incomplete geodesics are more appropriate for characterizing singularities than curvature scalars.

In a future communication we shall investigate how the exotic singularities presented here connect with the global picture and the underlying causal structure. It is our hope that a better understanding of the transformation will make it possible to design new solutions of Maxwell equations in curved spacetimes with nontrivial topological aspects.

\section{Acknowledgements}
EB is partially supported by \textit{Conselho Nacional de Desenvolvimento Cient\'ifico e Tecnol\'ogico} (CNPq) under Grant No.\ 305217/2022-4. RF is supported by \textit{Coordenação de Aperfeiçoamento de Pessoal de Nível Superior} (CAPES) under Grant No.\ 88887.671331/2022-00. RF would like to thank the hospitality of \textit{Scuola Superiore Meridionale} (Naples/Italy) while this work was being done.
\section*{References}
\bibliography{refs.bib}

\providecommand{\newblock}{}
\begin{thebibliography}{10}
\expandafter\ifx\csname url\endcsname\relax
  \def\url#1{{\tt #1}}\fi
\expandafter\ifx\csname urlprefix\endcsname\relax\def\urlprefix{URL }\fi
\providecommand{\eprint}[2][]{\url{#2}}

\bibitem{Penrose}
Penrose R 1965 {\em Physical Review Letters\/} {\bf 14.3} 57

\bibitem{Hawking}
Hawking S~W 1966 {\em Physical Review Letters\/} {\bf 17} 444

\bibitem{hawking1970singularities}
Hawking S~W and Penrose R 1970 {\em Proceedings of the Royal Society of London.
  A. Mathematical and Physical Sciences\/} {\bf 314} 529--548

\bibitem{hawking2023large}
Hawking S~W and Ellis G~F 2023 {\em The large scale structure of space-time\/}
  (Cambridge university press)

\bibitem{senovilla1998singularity}
Senovilla J~M 1998 {\em General Relativity and Gravitation\/} {\bf 30} 701--848

\bibitem{senovilla2011singularity}
Senovilla J~M 2011 Singularity theorems in general relativity: Achievements and
  open questions {\em Einstein and the changing worldviews of physics\/}
  (Springer) pp 305--316

\bibitem{senovilla20151965}
Senovilla J~M and Garfinkle D 2015 {\em Classical and Quantum Gravity\/} {\bf
  32} 124008

\bibitem{ellis1977singular}
Ellis G~F and Schmidt B~G 1977 {\em General Relativity and Gravitation\/} {\bf
  8} 915--953

\bibitem{Gordon}
Gordon W 1923 {\em Ann. Phys. (Leipzig)\/} {\bf 72} 421–456

\bibitem{kerr-schild}
Kerr R and Schild A 2009 {\em General Relativity and Gravitation\/} {\bf 41}
  2485--2499

\bibitem{beken1}
Bekenstein J~D 1993 New gravitational theorie as alternatives to dark matter
  {\em The Sixth Marcel Grossmann Meeting\/} ed Sato H and Nakamura T (WORLD
  SCIENTIFIC)
  \urlprefix\url{https://www.worldscientific.com/doi/abs/10.1142/1644}

\bibitem{beken2}
Bekenstein J 1993 {\em Physical Review D\/} {\bf 48} 3641

\bibitem{Nov_Vis_Vol}
Novello M, Visser M and Volovik G 2002 {\em Artificial Black Holes\/} (WORLD
  SCIENTIFIC)
  \urlprefix\url{https://www.worldscientific.com/doi/abs/10.1142/4861}

\bibitem{Barcelo2005}
Barcel\'o C, Liberati S and Visser M 2005 {\em Living Reviews in Relativity\/}
  {\bf 8}(1) 12

\bibitem{beken_mond04}
Bekenstein J 2004 {\em Physical Review D\/} {\bf 70} 083509 [Erratum-ibid. D
  {\bf 71} 069901 (2005)]

\bibitem{magueijo04}
Magueijo J and Smolin L 2004 {\em Classical and Quantum Gravity\/} {\bf 21}
  1725

\bibitem{Kaloper04}
Kaloper N 2004 {\em Physics Letters B\/} {\bf 583} 1--13

\bibitem{milgrom09}
Milgrom M 2009 {\em Physical Review D\/} {\bf 80} 123536

\bibitem{clifton12}
Clifton T, Ferreira P, Padilla A and Skordis C 2012 {\em Physics Reports\/}
  {\bf 513} 1

\bibitem{mota12}
Koivisto T, Mota D and Zumalacarregui M 2012 {\em Physical Review Letters\/}
  {\bf 109} 241102

\bibitem{Novello12}
Novello M and Bittencourt E 2012 {\em Phys. Rev. D\/} {\bf 86}(12) 124024

\bibitem{Novello13}
Novello M and Bittencourt E 2013 {\em General Relativity and Gravitation\/}
  {\bf 45}(5) 1005

\bibitem{scalartheory13}
Novello M, Bittencourt E, Moschella U, Goulart E, Salim J and Toniato J 2013
  {\em Journal of Cosmology and Astroparticle Physics\/} {\bf 06} 014

\bibitem{dario13}
Bettoni D and Liberati S 2013 {\em Physical Review D\/} {\bf 88} 084020

\bibitem{Bit14}
Bittencourt E, Faci S and Novello M 2014 {\em International Journal of Modern
  Physics A\/} {\bf 29} 1450145

\bibitem{Nov_Bit14}
Novello M and Bittencourt E 2014 {\em International Journal of Modern Physics
  A\/} {\bf 29} 1450075

\bibitem{rua14}
Deruelle N and Rua J 2014 {\em Journal of Cosmology and Astroparticle
  Physics\/} {\bf 09} 002

\bibitem{sakstein14}
Sakstein J 2014 {\em Journal of Cosmology and Astroparticle Physics\/} {\bf 09}
  012

\bibitem{Arroja15}
Arroja F, Bartolo N, Karmakar P and Matarrese S 2015 {\em Journal of Cosmology
  and Astroparticle Physics\/} {\bf 2015} 051

\bibitem{Ip15}
Ip H~Y, Sakstein J and Schmidt F 2015 {\em Journal of Cosmology and
  Astroparticle Physics\/} {\bf 2015} 051

\bibitem{Sakstein15}
Sakstein J and Verner S 2015 {\em Phys. Rev. D\/} {\bf 92}(12) 123005

\bibitem{Yuan15}
Yuan F~F and Huang P 2015 {\em Physics Letters B\/} {\bf 744} 120--124 ISSN
  0370-2693

\bibitem{Carvalho16}
Carvalho G~G, Lobo I~P and Bittencourt E 2016 {\em Phys. Rev. D\/} {\bf 93}(4)
  044005 \urlprefix\url{https://link.aps.org/doi/10.1103/PhysRevD.93.044005}

\bibitem{Falciano12}
Falciano F~T and Goulart E 2012 {\em Classical and Quantum Gravity\/} {\bf 29}
  085011 \urlprefix\url{https://dx.doi.org/10.1088/0264-9381/29/8/085011}

\bibitem{bitt15}
Bittencourt E, Lobo I~P and Carvalho G~G 2015 {\em Classical and Quantum
  Gravity\/} {\bf 32} 185016

\bibitem{Goulart13}
Goulart E and Falciano F~T 2013 {\em Classical and Quantum Gravity\/} {\bf 30}
  155020 \urlprefix\url{https://dx.doi.org/10.1088/0264-9381/30/15/155020}

\bibitem{Goulart21}
Érico Goulart and Bittencourt E 2021 {\em Classical and Quantum Gravity\/}
  {\bf 38} 145029 \urlprefix\url{https://dx.doi.org/10.1088/1361-6382/ac08a9}

\bibitem{harte2017metric}
Harte A~I 2017 {\em Physical Review Letters\/} {\bf 118} 141101

\bibitem{Syn}
Synge J~L 1965 {\em Relativity: the special theory\/} (North-Holland)

\bibitem{Lan}
Landau L~D 2013 {\em The classical theory of fields. Vol. 2\/} (Elsevier
  Science)

\bibitem{Sch}
Schouten J~A 2013 {\em Ricci-calculus: an introduction to tensor analysis and
  its geometrical applications\/} (Springer Science and Business Media)

\bibitem{Wal}
Wald R~M 2010 {\em General relativity\/} (University of Chicago Press)

\bibitem{misner1967taub}
Misner C~W 1967 {\em Relativity theory and astrophysics\/} {\bf 1} 160

\bibitem{Carr1965}
Carr R~E 1965 Dipolar coordinates Tech. rep. California Institute of Technology
  \urlprefix\url{https://core.ac.uk/reader/85255358}

\bibitem{gautreau1967directional}
Gautreau R and Anderson J~L 1967 {\em Physics Letters A\/} {\bf 25} 291--292

\bibitem{taylor2005unravelling}
Taylor J 2005 {\em Classical and Quantum Gravity\/} {\bf 22} 4961

\end{thebibliography}
\end{document}